\documentclass[9pt]{article}
\usepackage{spconf,amsmath,graphicx}
\usepackage[section]{placeins}
\usepackage{url}
\usepackage{relsize}
\usepackage{hyperref}
\usepackage{epstopdf}
\usepackage{enumerate}
\usepackage{algorithmicx}
\usepackage{algorithm}
\usepackage{amsmath}
\usepackage[noend]{algpseudocode}
\usepackage{float}
\usepackage{color}
\usepackage{makeidx}
\usepackage{bbm}
\usepackage{graphicx}
\usepackage{lipsum}
\usepackage{subcaption}
\usepackage[section]{placeins}
\usepackage{comment}

\usepackage{xspace}
\newcommand{\eg}{\textit{e.g.},\xspace}
\newcommand{\ie}{\textit{i.e.},\xspace}
\usepackage{setspace}
\usepackage{lipsum}
\usepackage{adjustbox}
\usepackage{colortbl}
\usepackage{hhline}

\usepackage{color}
\usepackage[dvipsnames]{xcolor}

\usepackage{multicol}
\definecolor{Gray}{gray}{0.77}
\definecolor{Grayy}{gray}{0.87}

\usepackage{lipsum}
\let\OLDthebibliography\thebibliography
\renewcommand\thebibliography[1]{
  \OLDthebibliography{#1}
  \setlength{\parskip}{0pt}
  \setlength{\itemsep}{0pt plus 0.3ex}
}

\title{The First Indoor Pathloss Radio Map Prediction Challenge}
\name{Stefanos Bakirtzis$^{\parallel, *, \dagger, \$}$  \quad \c{C}a\u{g}kan Yapar$^{\parallel, \ddagger }$ \quad Kehai Qiu$^{*, {\P}, \$}$ \quad Ian Wassell$^{*}$ \quad Jie Zhang$^{\S, \#}$
}

\address{
    $^{*}$University of Cambridge,    $^{\ddagger}$Technical University of Berlin\\
    $^{\dagger}$Telefonica Research, $^{\P}$Brunel University London,  $^{\S}$Ranplan Wireless, 
    $^{\#}$University of Sheffield
}
\newcounter{CagkanCounter}

\begin{document}
%

\maketitle
\def\thefootnote{$\parallel$}\footnotetext{These authors contributed equally to this work.}\def\thefootnote{\arabic{footnote}}
\def\thefootnote{\$}\footnotetext{At the time of submission the authors were also with Ranplan Wireless. Stefanos Bakirtzis' work is supported by the  EPSRC IAA Strategic Impact Partnership Fund and the Foundation for Education and European Culture.}\def\thefootnote{\arabic{footnote}}
\vspace{-3.2mm}
\begin{abstract}
\vspace{-3.2mm}
To encourage further research and to facilitate fair comparisons in the development of deep learning-based radio propagation models, in the less explored case of directional radio signal emissions in indoor propagation environments, we have launched the  ICASSP 2025 First Indoor Pathloss Radio Map Prediction Challenge.  This overview paper describes the indoor path loss prediction problem, the datasets used, the Challenge tasks, and the evaluation methodology. Finally, the results of the Challenge and a summary of the submitted methods are presented.  

\end{abstract}
\vspace{-2.2mm}
%
\vspace{-6.2mm}
\section{Introduction} \label{sec:Intro}
\vspace{-4.5mm}

 Radio propagation models emulate various phenomena, including fading, path loss, shadowing, and interference, which influence the received signal strength at the user equipment. Consequently, they can streamline various wireless network applications, such as coverage prediction,  fingerprint-based localization,  and network planning \cite{WCM_paper}. A widely adopted measure for quantifying large-scale fading is \emph{path loss} (PL), which indicates the power attenuation of the emitted signals, due to free-space losses or because of the electromagnetic wave interaction with obstacles in the propagation environment, \eg reflection,  refraction, or diffraction.

    Recent works exploit advances in deep learning (DL) to develop data-driven methods \cite{RadioUNetTWC, EMDeepRay} that can learn from observation of real-world or simulated data. These models can yield commensurate accuracy with deterministic propagation models, accompanied by impressive computational efficiency due to their straightforward graph processing unit parallelization. However, while previous research has focused mostly on DL-based PL inference for outdoor urban environments and isotropic transmissions,  due to the wide deployment of indoor wireless networks (IWNs) in fifth-generation and beyond networks, it is necessary to develop models tailored for indoor environments. 
    In that case, accurate indoor radio map estimation requires accounting for the larger variety of construction materials and their electromagnetic properties. Likewise, it is imperative to consider the impact of the antenna radiation pattern in the propagation of radio waves.\looseness=-1

    Motivated by the fruitful developments induced by the previous Pathloss Radio Map Challenge at ICASSP 2023 \cite{challenge23}, we shared a simulated indoor PL radio map dataset to accelerate the progress in PL radio map prediction for IWNs.
    



\vspace{-5.6mm}
\section{Datasets}\label{sec:Datasets}
\vspace{-4.5mm}
\subsection{Training Dataset}\label{sec:3DDataset}
\vspace{-2.2mm}

The \emph{Indoor Radio Map Dataset}  \cite{DataSetAlt} comprises PL radio maps generated via the \emph{Ranplan 
Wireless}\footnote{\url{https://www.ranplanwireless.com/}} 
 ray tracing 
software in different settings.    Specifically, the dataset includes PL radio maps from 25 indoor environments of different sizes, and complexity, comprising various construction material types (\eg concrete, plasterboard, wood, glass, metal),  3 frequency bands (868 MHz, 1.8 GHz, and 3.5 GHz), and 5 antenna radiation patterns including the case of isotropic antennas.

For all the simulations the transmitter (Tx) height is set at 1.5 m above the floor, while we are sampling the PL distribution at a receiving plane at the same height.  The spatial resolution is set to 0.25 m, \ie we sample PL every 0.25 meters, and we trace the ray path up to 8 reflections, 10 transmissions, and 2 diffractions.  The input tensors are represented as RGB images, where the first two channels indicate the (i) normal incidence reflectance (0 for air) and (ii) incidence transmittance (0 for air) absolute values at each point of the grid, measured in dB, as well as (iii) the physical distance between the Tx and each point of the grid. In addition, for each sample, we provide the Tx location, operating frequency, and antenna radiation pattern. Finally, the target PL radio maps are represented as grayscale images where each point of the image denotes the PL at that point. The minimum and maximum PL values in the dataset were 13 dB and 160 dB, respectively. 

\vspace{-5.2mm}
\subsection{Test Dataset}
\vspace{-2.2mm}

To evaluate the candidate solutions we generated an additional test dataset that was not published, which includes samples from 5 geometries, a known and unknown frequency band (868 MHz and 2,4 GHz), and two antenna radiation patterns not included in the training dataset. The samples for the test dataset were created through the same procedure followed for the {Indoor Radio Map Dataset}.

\vspace{-4.9mm}
\section{The Challenge Tasks}
\vspace{-3.4mm}

\begin{table*}[t!]
\vspace{-2.2mm}
		\caption{\small Accuracies of the submitted methods on the test set. } \label{table:RMSETable}
		\vspace{-3.6mm}	
	\renewcommand{\arraystretch}{1}

	\centering
	\scalebox{1.05}{
	\resizebox{2.0\columnwidth}{!}{
	\begin{tabular}{c|c c c c c | c c c c c }
		\hline	
		\cellcolor{blue!15} {\bfseries  Method }& \cellcolor{Grayy}\bfseries SIP2NET \cite{SIP2NET}  &\cellcolor{Grayy} IPP-NET \cite{IPP-NET}&\cellcolor{Grayy} TerRaIn \cite{Split_U_Net}&\cellcolor{Grayy} TransPathNet \cite{TransPathNet} &\cellcolor{Grayy} ResUNet \cite{ResUnet}&\cellcolor{Gray} Team 6  &\cellcolor{Gray} Team 7 \cite{VietDanielTemp}&\cellcolor{Gray}Team 8 \cite{8Alt}&\cellcolor{Gray} Team 9 &\cellcolor{Gray} Team 10 \cite{YuukiTemp}   \\
		\hline
	
		  \begin{tabular}{@{}c@{}} Weighted  RMSE (dB) \end{tabular}       & 9.41 & 9.50  &  10.32  &  10.39 & 11.09  &  11.30 &  13.25  &  13.69  &  13.73  & 14.47 \\

		\hline

	\end{tabular}
	}}
\vspace{-7.8mm}
\end{table*}

The Challenge\footnote{\url{https://IndoorRadioMapChallenge.GitHub.io/}} consists of three supervised learning tasks, aiming at exploring how well a data-driven model can generalize to different IWN propagation environments and settings. Specifically, we probe the generalizability of the models  ($i$) only for unknown indoor environment scenes (Task 1), ($ii$) both for new building layouts and frequency bands (Task 2), and ($iii$) simultaneously over new building layouts, frequency bands, and antenna radiation patterns (Task 3).

\vspace{-6.5mm}

\section{Challenge Results}

\vspace{-4.28mm}

\subsection{Evaluation Methodology}

\vspace{-3.2mm}

The participants submitted the radio maps generated by their DL-based models based on each Task test set input data (which was sent to them without the ground truth). Alongside, the participants submitted the code to train the models and run their evaluations. We assessed the fidelity of each model by computing the root mean square error (RMSE) for each Task as defined in  Equation (7)  of \cite{EMDeepRay}. The final score for each team was calculated as the weighted average of the three Task RMSEs, with the weights set to 0.3, 0.3, and 0.4, for Tasks 1, 2, and 3, respectively.

\begin{figure}[t]
\vspace{-1.1mm} 

\hspace{  50 mm} 
\subfloat{\includegraphics[width=0.45\columnwidth]{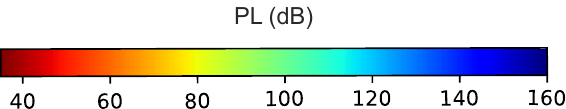}}
\vspace{0.1mm}

\centering
\subfloat{\includegraphics[width=0.175\columnwidth]{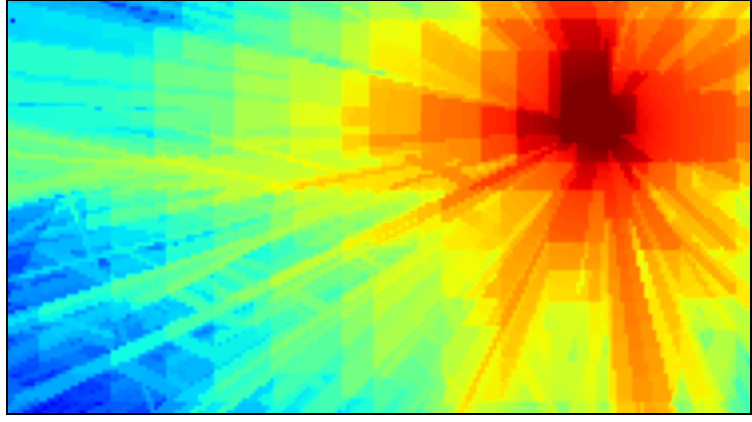}}
\subfloat{\includegraphics[width=0.175\columnwidth]{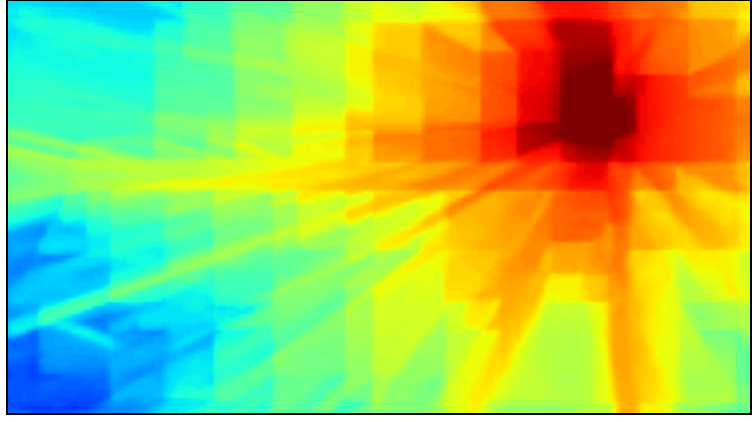}}
\subfloat{\includegraphics[width=0.175\columnwidth]{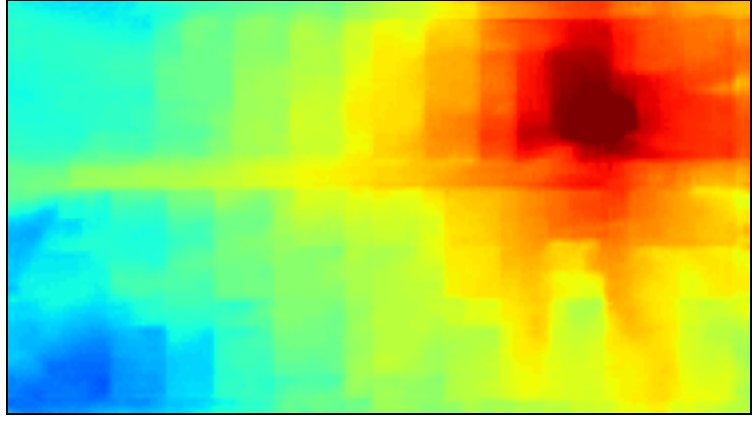}}
\subfloat{\includegraphics[width=0.175\columnwidth]{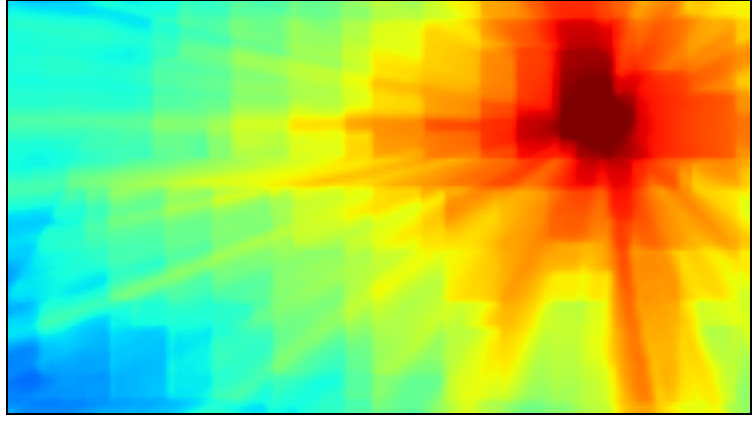}}
\subfloat{\includegraphics[width=0.175\columnwidth]{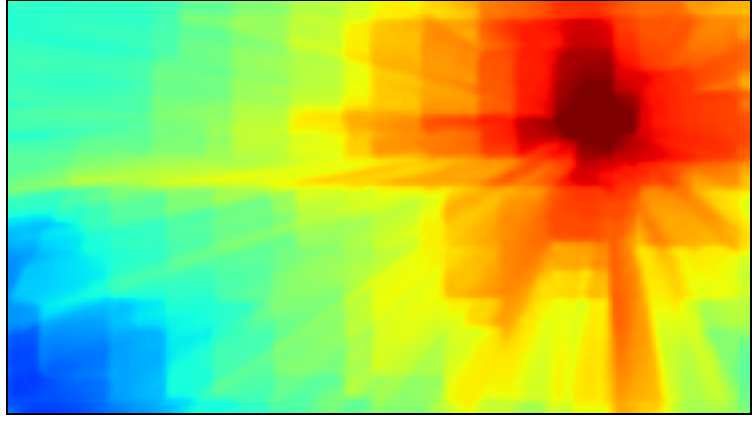}}
\subfloat{\includegraphics[width=0.175\columnwidth]{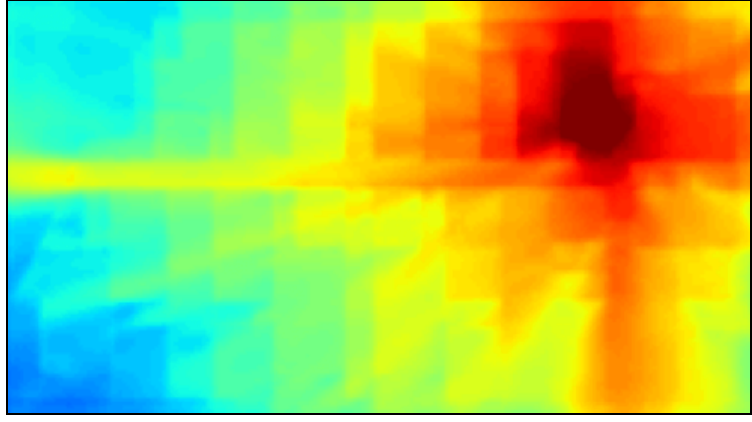}}

\captionsetup[sub]{font=scriptsize}
\captionsetup[subfigure]{skip=1pt}
\setcounter{subfigure}{0}
\subfloat[GT]{\includegraphics[width=0.175\columnwidth]{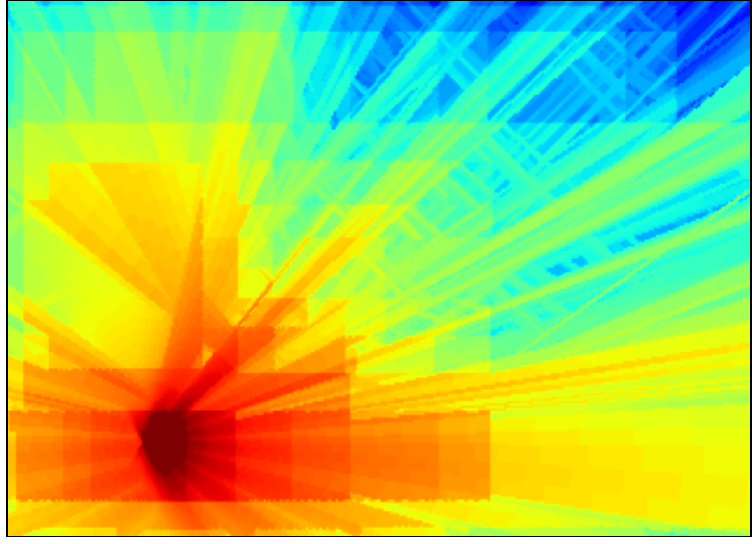}}
\subfloat[SIP2NET]{\includegraphics[width=0.175\columnwidth]{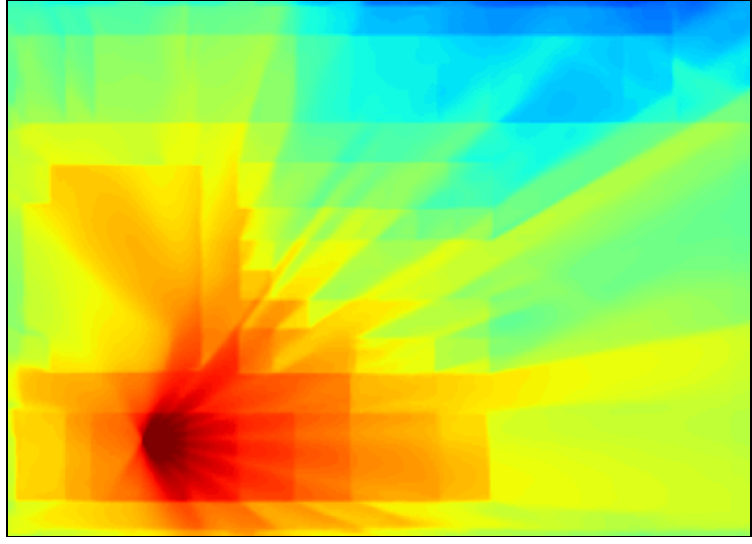}}
\subfloat[IPP-NET]{\includegraphics[width=0.175\columnwidth]{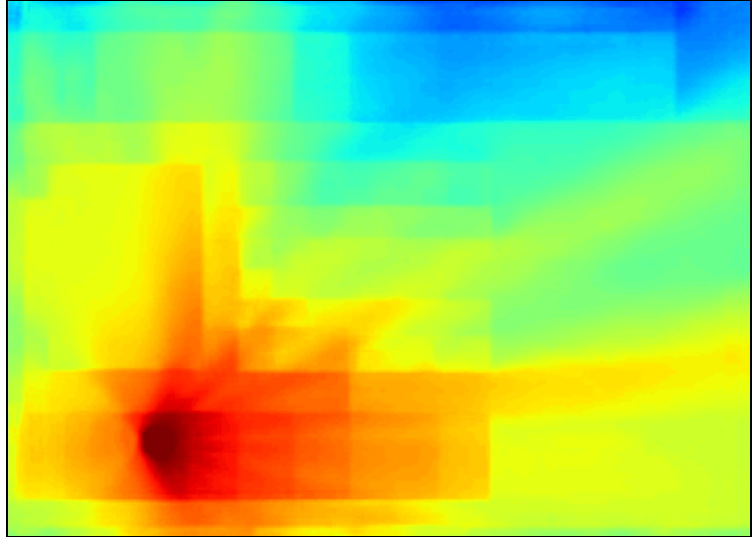}}
\subfloat[TerRaIn]{\includegraphics[width=0.175\columnwidth]{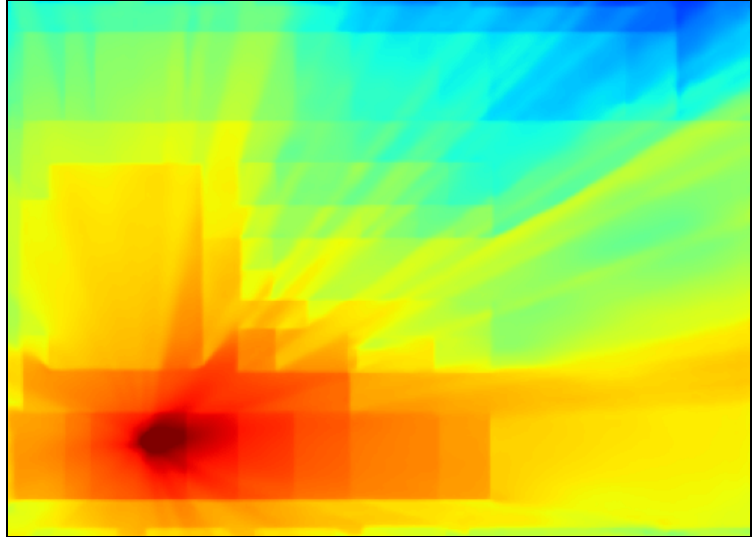}}
\subfloat[TransPath]{\includegraphics[width=0.175\columnwidth]{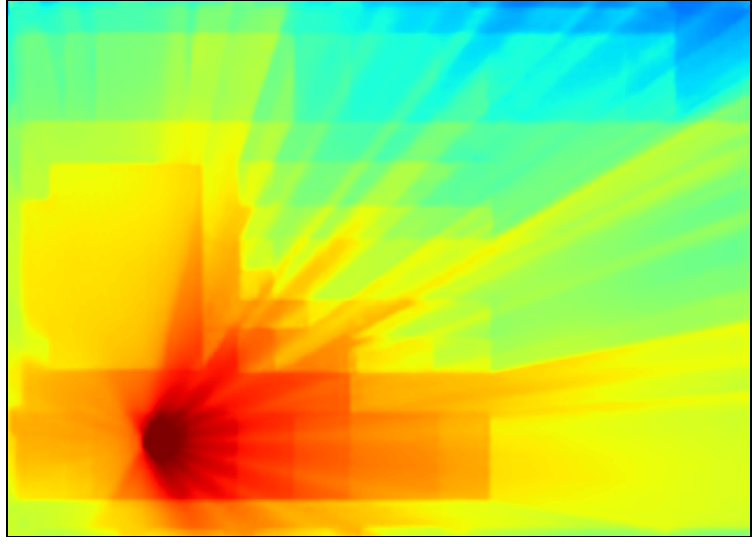}}
\subfloat[ResUNet]{\includegraphics[width=0.175\columnwidth]{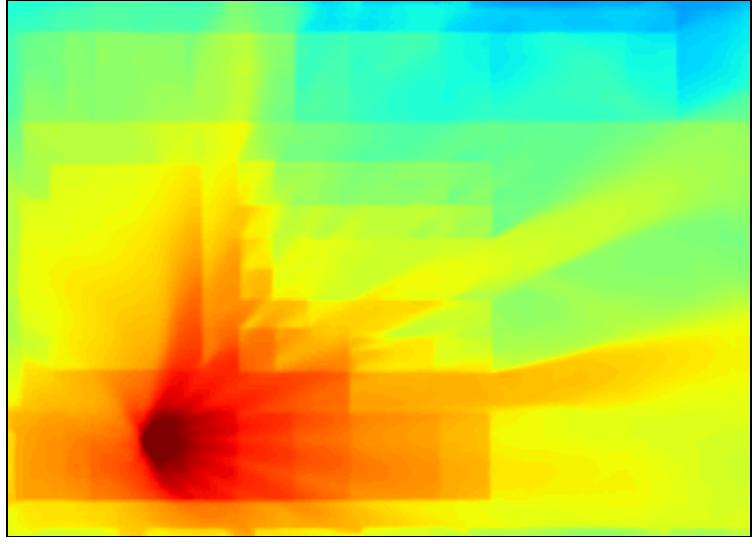}} 
\vspace{-3.8mm}
\caption{Visual comparison of PL radio maps.}
\label{fig:best_bacelli_model_prediction}
\vspace{-8mm}
\end{figure}
 
\vspace{-5.2mm}

\subsection{Submitted Methods and Evaluation Results}
\vspace{-2.2mm} 
 
 Table \ref{table:RMSETable} summarizes the final score attained by the top 10 ranked teams, while  Fig. \ref{fig:best_bacelli_model_prediction}  depicts some indicative PL radio maps generated by the top 5 teams comparing them to the ground truth (GT) maps.  All participants used additional physics-inspired input features, such as the free space path loss (FSPL), and data augmentation techniques to varying degrees, again proving their previously observed importance in achieving high-accuracy radio map predictions \cite{challenge23}. The participants reported average run-times of about 20 ms (on different system setups, see the participants' papers for details), thereby complying with the challenge requirements.

 The best-performing model, SIP2NET \cite{SIP2NET}, employed a U-Net-like architecture using asymmetric convolutions in the encoder and decoder to enhance detail generation and Atrous Spatial Pyramid Pooling (ASPP) in the bottleneck layer to extract multi-scale features. The authors used several loss functions \textcolor{black}{including a generative adversarial loss} to increase the accuracy, attaining a final weighted RMSE score of 9.41 dB. 
 
 IPP-NET \cite{IPP-NET} also builds on a U-Net architecture and is equipped with five consecutive dilated convolution layers and a standard convolution. In addition to the input features provided in the dataset (cf. Sec. 2.1), IPP-Net receives as input a PL radio map calculated by a modified 3GPP Indoor Hotspot PL model, corresponding to a Bresenham's line algorithm-based non-line-of-sight map with learnable weights. 
 
 Instead of using a single encoder-decoder, the authors of \cite{Split_U_Net} used two parallel encoder-decoder streams: one with stacked dilated convolutions for multi-scale feature extraction and one with standard convolutions to process sparse data. 
 
TransPathNet \cite{TransPathNet} is composed of a transformer-based feature extraction encoder in conjunction with a multiscale convolutional attention decoding module, employing several additional input features and a coarse-to-fine two-stage strategy.  
 
 In \cite{ResUnet}, the authors used a ResUNet architecture incorporating dual-stream processing for frequency adaptation, diverse data augmentation techniques, and employed several physics-informed input features and a multitude of loss terms. 
 
 Runner-up teams also presented interesting approaches; for example, the authors of \cite{VietDanielTemp} used a PL radio map that is based on FSPL, transmission losses, and radiation pattern as an input feature; while the authors of \cite{8Alt} proposed a Vision Transformer (ViT)-based method employing extensive data augmentation techniques and using weights of a pre-trained ViT; and \cite{YuukiTemp} introduced the application of MST++ for the task of radio map estimation.


Finally, we note that we consistently observed discrepancies in errors across different buildings. The larger buildings exhibited the highest errors, especially in areas far from the Tx where the PL reached its maximum value (160 dB).  Another major source of large errors was cases where the Tx was almost on top of the walls, resulting in a large error offset for the entire map; this effect was more pronounced for buildings with dense wall layouts.  

\vspace{-4.3mm}


\bibliographystyle{IEEEbib}
 
{\scriptsize
\bibliography{pub}
}


\end{document}